# The Origination and Diagnostics of Uncaptured Beam in the Tevatron and Its Control by Electron Lenses

Xiao-Long Zhang[1], Kip Bishofberger[2], Vsevolod Kamerdzhiev[1], Valery Lebedev[1],

Vladimir Shiltsev[1], Randy Thurman-Keup[1], Alvin Tollestrup[1]

[1]Fermi National Accelerator Laboratory, PO Box 500, Batavia, IL 60510, USA

[2]Los Alamos National Laboratory, Los Alamos, NM 87545, USA

*Abstract*

In the Collider Run II, the Tevatron operates with 36 high intensity bunches of 980 GeV protons and antiprotons. Particles not captured by the Tevatron RF system pose a threat to quench the superconducting magnet during acceleration or at beam abort. We describe the main mechanisms for the origination of this uncaptured beam, and present measurements of its main parameters by means of a newly developed diagnostics system. The Tevatron Electron Lens is effectively used in the Collider Run II operation to remove uncaptured beam and keep its intensity in the abort gaps at a safe level.

# 1. INTRODUCTION

The Tevatron is a 6.3 km long circular collider operating with 36 proton and 36 antiproton bunches at a beam energy of 980 GeV. The main parameters of the Collider are given in Table 1 and a description of operation can be found in Ref. [1]. The 36 bunches in each beam are grouped in three trains of 12 bunches with a bunch spacing of 396 ns, which is equal to 21 RF buckets. The bunch trains are separated by three 2.52 μs long abort gaps. High intensity proton bunches are generated by coalescing several (usually seven) smaller bunches in Fermilab's Main Injector (MI) at 150 GeV before being injected into the Tevatron. The injection process takes about half an hour, then both beams are being accelerated to 980 GeV in about 90 seconds, and stay at the flat-top energy for the rest of the high-energy physics (HEP) store.

Coalescing in the MI typically leaves a few percent of the beam particles outside RF buckets. These particles are transferred together with the main bunches. In addition, single intra-beam scattering (known as Touschek effect [2]) and diffusion due to multiple intra-beam scattering (IBS), as well as the phase and amplitude noise of the RF voltage, drive particles out of the RF buckets. It is exacerbated by the fact that after coalescing and injection, 95% of particles cover almost the entire RF bucket area. To prevent longitudinal instabilities, which can blow-up the bunch length and drive particles out of the RF buckets, a longitudinal bunch-by-bunch feedback system has been installed [3]. The uncaptured beam is lost at the very beginning of the Tevatron energy ramp. These particles are out-of-sync with the Tevatron RF accelerating system, so they do not gain energy and quickly (< 1 sec) spiral radially into the closest horizontal aperture. If the number of particles in the uncaptured beam is too large, the corresponding energy deposition results in a quench (loss of superconductivity) of the SC magnets and, consequently, terminates the high-energy physics store. At the injection energy, an instant loss of uncaptured beam equal to 3-7% of the total intensity can lead to a quench depending on the spatial distribution of the losses around the machine circumference.

At the top energy, uncaptured beam generation is mostly due to the IBS and RF noise while infrequent occurrences of the longitudinal instabilities or trips of the RF power amplifiers can contribute large spills of particles to the uncaptured beam. Uncaptured beam particles are outside of the RF buckets, and therefore, move longitudinally relative to the main bunches. Contrary to the situation at the injection energy of 150 GeV, when synchrotron radiation (SR) losses are practically negligible, 980 GeV protons and antiprotons lose about 9 eV/turn due to the SR. For uncaptured beam particles, this energy loss is not being replenished by the RF system, so they slowly spiral radially inward and die on the collimators, which determine the tightest aperture in the Tevatron during collisions. The typical time for an uncaptured beam particle to reach the collimator is about 20 minutes.

Table 1. Tevatron Collider Run II Parameters

| Parameter | Symbol | Value | Units |
|---|---|---|---|
| **Beam Energy** | $E$ | 980 | GeV |
| **Peak luminosity** | $L$ | 2.92 | $10^{32}$ cm$^{-2}$s$^{-1}$ |
| **Circumference** | $C$ | 6280 | m |
| **Number of bunches** | $N_b$ | 36 | |
| **Protons/bunch** | $N_p$ | 250-300 | $10^9$ |
| **Antiprotons/bunch** | $N_a$ | 40-100 | $10^9$ |
| **RF voltage** | $V_{RF}$ | 1 | MV |
| **RF frequency** | $f_{RF}$ | 53.1 | MHz |
| **RF harmonic number** | $h$ | 1113 | |
| **Bunch spacing** | $t_b$ | 396 | ns |
| **Momentum compaction factor** | $\eta$ | 0.0028 | |
| **RF bucket area** | $S_{RF}$ | 4.4 @150GeV<br>11.0@980GeV | eV s |
| **Longitudinal emittance at the start of store, 95%** | $\varepsilon_p$ & $\varepsilon_a$ | 3 - 4 | eV s |
| **Proton/antiproton bunch length** | $\sigma_s$ | 3.0 @150GeV<br>1.7 @980 GeV | ns |
| **Energy loss per turn due to SR** | $eV_{SR}$ | 9.5 | eV |
| **Synchrotron frequency** | $f_s$ | 87 @150GeV<br>35 @980 GeV | Hz |
| **Synchrotron tune** | $\nu_s$ | 1.7 @150GeV<br>0.7 @980 GeV | $10^{-3}$ |

The total uncaptured beam intensity is a product of the rate at which particles leak out of the main bunches and the time required for them to leave the machine. If SR is the only energy loss mechanism, then during a typical HEP store as many as $60\times10^9$ particles could be accumulated in the uncaptured beam. Since uncaptured beam particles are distributed all around the circumference, those that are in the abort gaps between the bunch trains will be deposited into nearby magnets and other places that limit machine acceptance whenever the abort kicker fires. The resulting quenches were of great concern in Dec.2001-Feb.2002 as they greatly affected the collider operation. Note that for protons, which are the major contributor to the uncaptured beam, the vacuum chamber in the vicinity of the CDF detector is one of the tightest apertures encountered by the beam and loss around the detector poses a great threat due to radiation damage of detector components. Use of the Tevatron Electron Lens (TEL) [4] reduced the uncaptured beam removal time from 20 min to about 2 min thereby significantly reducing its intensity and as a result, the quenches on abort due to the uncaptured beam disappeared completely. In the following sections we will look into the dynamics of the uncaptured beam generation, discuss diagnostic tools used for monitoring the uncaptured beam parameters, and describe the basics of the TEL operation in the regime of uncaptured beam cleaning.

## 2. UNCAPTURED BEAM FORMATION

*2.1 Beam dynamics of the longitudinal phase space*

In the case of single harmonic RF, a particle phase trajectory in the longitudinal phase space (see Figure 1) is described by the following equation [5]:

$$\left(\frac{\Delta p}{p_0}\right)^2 = \frac{2v_s^2}{\eta^2 h^2}\left(\cos(\phi) - \cos(\phi_m) + (\phi - \phi_m)\sin\varphi_0\right), \quad (1)$$

where ($\Delta p/p_0$) is the relative particle momentum deviation, $\eta$ is the slip factor, $h$ is the harmonic number, $v_s$ is the synchrotron tune (see Table 1), $\phi$ is the RF phase, $\phi_0$ is the accelerating phase and $\phi_m$ determines the boundary of phase space trajectory. In the stationary state $\phi_0$ is determined by particle energy loss due to synchrotron radiation $eV_{SR}$: $\sin\varphi_0 = V_{SR}/V_{RF}$. The SR radiation damping is neglected in Eq.(1) as its time is much longer than the store duration.

The outermost orbit, called the separatrix, determines the boundary of RF bucket:

$$\left(\frac{\Delta p}{p_0}\right)^2 = \frac{2v_s^2}{\eta^2 h^2}\left(\cos(\phi) + \cos(\phi_0) + (\phi - \pi + \phi_0)\sin\varphi_0\right). \quad (2)$$

If $\phi_0 \ll 1$, the separatrix boundaries in RF phase are described by the following equations

$$\phi_1 \approx -\pi + \sqrt{4\pi\phi_0} - \frac{\phi_0}{2},$$
$$\phi_2 = \pi - \phi_0. \quad (3)$$

Figure 1 presents the corresponding phase space trajectories for $\phi_0$=0.15. In the case of the Tevatron during collisions $\phi_0 \approx 10^{-11}$, $\phi_1 \approx -\pi + 10^{-5}$ and $\phi_2 \approx \pi - 10^{-11}$. Thus, the

Tevatron RF buckets are separated by a gap of ~$10^{-5}$ rad. A particle with initial momentum above the RF bucket boundary is decelerated by SR, but it will never penetrate the RF bucket. Instead it penetrates a gap between buckets to the lower momentum side and, finally, it is decelerated to the nearest apertures limiting the beam energy spread.

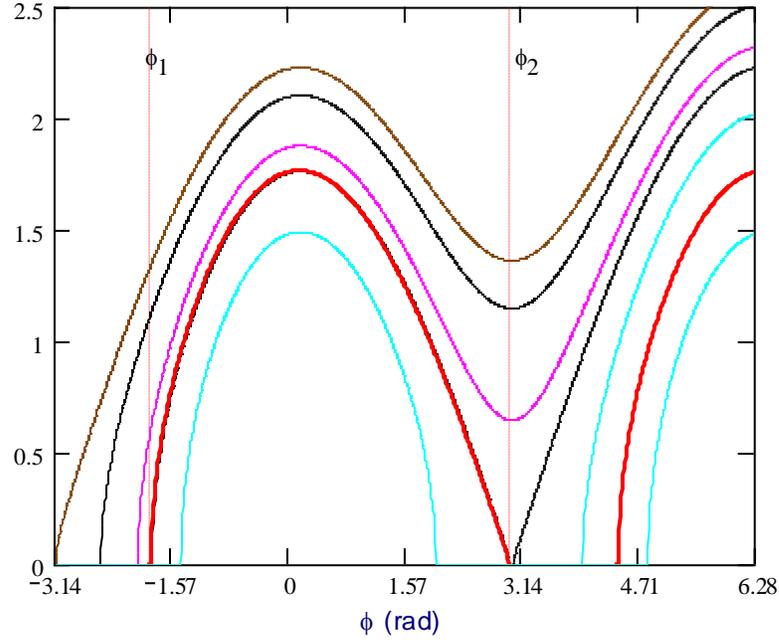

Figure 1. Upper half of phase space trajectories in the vicinity of the separatrix (red line) for $\phi_0$=0.15. Momentum spread (vertical axis) is presented in units of

$$(\eta h / v_s) \Delta p / p_0 .$$

*2.2 Longitudinal beam diffusion rate*

There are three major mechanisms creating uncaptured beam. They are the diffusion due to amplitude and phase RF noises[6], multiple intrabeam scattering

(IBS) and single intrabeam scattering (Touschek effect)[2]. Immediately after acceleration, the bunch occupies approximately 4.4 eVs of the longitudinal phase space, while the total RF bucket area is about 11 eV s. Therefore, there are no tails and single IBS is the only mechanism for particle loss. Shortly after injection, the diffusion due to IBS and RF noise creates tails in the distribution function and results in additional beam loss, which significantly exceeds the loss due to single IBS [7]. Therefore, we neglect the single IBS in the further analysis.

The diffusion equation in a sinusoidal longitudinal potential can be written in the following form [7]:

$$\frac{\partial f}{\partial t} = \frac{\partial}{\partial I}\left( I \frac{D(I)}{dE/dI} \frac{\partial f}{\partial I} \right) . \qquad (4)$$

Here $f=f(I, t)$ is the longitudinal distribution function, $D(I)$ is the diffusion coefficient, and $t$ is the time. $I$ and $E$ are the longitudinal action and the energy defined as:

$$I = \frac{1}{2\pi}\oint p d\phi, \quad E = \frac{p^2}{2} + \Omega_s^2(1-\cos\phi) \quad , \qquad (5)$$

where $p = \dot{\phi}$ is the canonical momentum, $\Omega_s = 2\pi\nu_s f_0$ is the synchrotron frequency. Figure 2 presents a numerical solution of this equation assuming i) constant diffusion, $D(I)=D_0$, as a zero-order approximation, ii) the initial distribution is a $\delta$-function, $f(I) = \delta(I)$, and iii) the boundary condition $f(I_{max}) = 0$ is met at the boundary on the RF bucket, $I_{max}= 8\Omega_s/\pi$. Figure 3 presents the corresponding beam intensity, rms momentum spread and rms bunch length. Initially, the bunch length and the momentum spread grow proportionally to $\sqrt{t}$ and the distribution function is close to the Gaussian distribution, $f(I,t) \propto (\Omega_s/Dt)\exp(-I\Omega_s/Dt)$. When the bunch length becomes comparable to the bucket length, the non-quadratic behavior of the potential results in the bunch-length growing faster than the momentum spread. Finally, the distribution function and, consequently, the bunch length and momentum spread approach their asymptotic values, and the intensity decays exponentially as $\sim \exp(-0.741 Dt/\Omega_s^2)$.

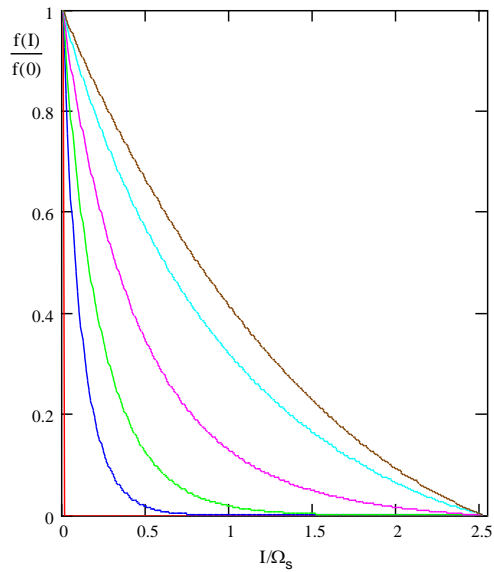

Figure 2: Dependence of the distribution function on time for $Dt/\Omega_s^2 = 0, 0.125, 0.25, 0.5, 1$ and asymptotic at $t \to \infty$.

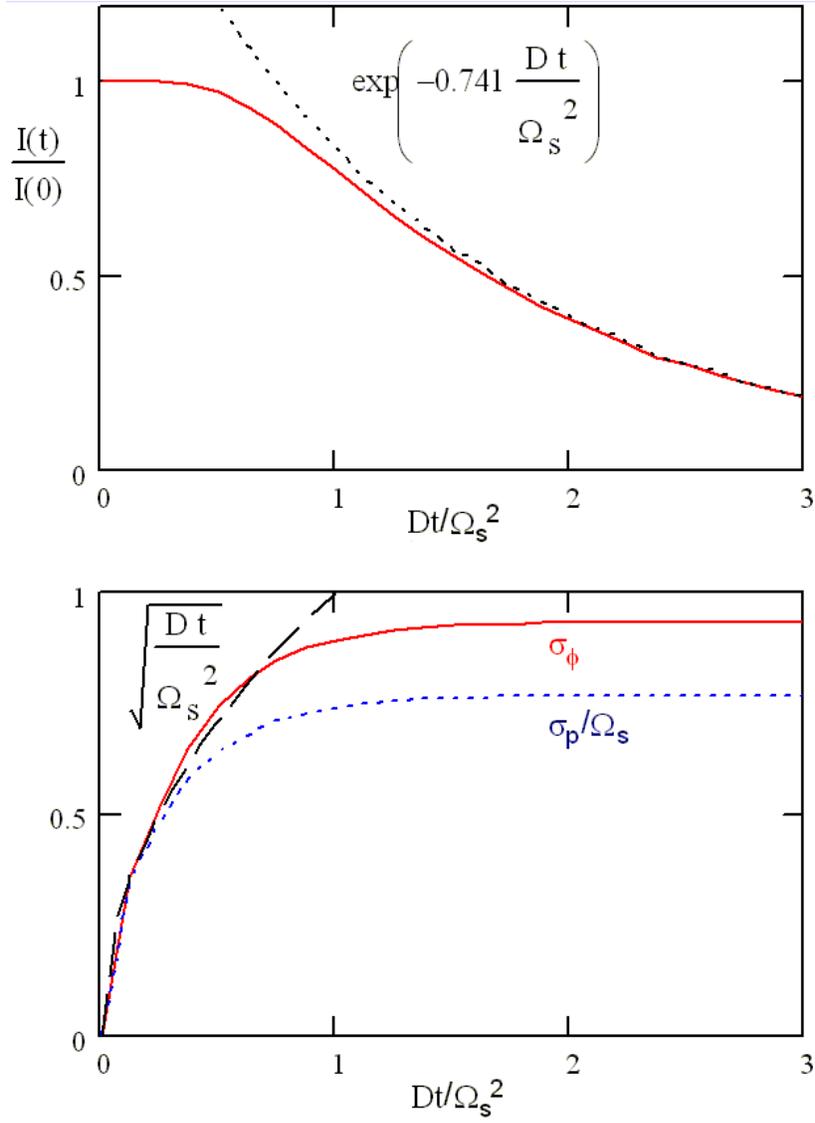

Figure 3: Time dependence of beam intensity (top) and rms bunch length and momentum spread (bottom).

The results of this simple model with constant diffusion and $f(I) = \delta(I)$ fit the evolution of the Tevatron bunch parameters and luminosity fairly well (within 10%) for the stores when the beam-beam effects are weak and the IBS effects dominate.

For the Tevatron collider parameters, the longitudinal energy spread in the beam rest frame is significantly smaller than the transverse ones (the ratio of the longitudinal kinetic energy to transverse kinetic energy is about 0.004 at the collision energy and about 0.02 at the injection energy). In this case, simplified IBS formulas can be used, e.g., for the momentum spread growth rate one gets [7]:

$$\frac{d(\sigma_{\Delta p/p}^2)}{dt} = \frac{r_p^2 cN}{4\sqrt{2}\gamma^3\beta^3\sigma_s}\left\langle\frac{\Xi_s(\theta_x,\theta_y)}{\sqrt{\theta_x^2+\theta_y^2}}\frac{L_C}{\sigma_x\sigma_y}\right\rangle_s, \qquad (6)$$

where $\gamma$ and $\beta$ are the relativistic factors, $r_p$ is the classical proton radius, $\sigma_{\Delta p/p}$ is the rms momentum spread, $\sigma_s$ is the rms bunch length,

$$\sigma_x = \sqrt{\varepsilon_x\beta_x + D_x^2\sigma_{\Delta p/p}^2}, \qquad \sigma_y = \sqrt{\varepsilon_y\beta_y},$$

$$\theta_x = \sqrt{\frac{\varepsilon_x}{\beta_x}\left(1+\frac{(D_x'\beta_x+\alpha_x D_x)^2\sigma_{\Delta p/p}^2}{\varepsilon_x\beta_x+D_x^2\sigma_{\Delta p/p}^2}\right)}, \qquad \theta_y = \sqrt{\varepsilon_y/\beta_y}, \qquad (7)$$

are the rms beam sizes and the local angular spreads, $\beta_x$, $\beta_y$, $\alpha_x$ and $\alpha_y$, are beta- and alpha-functions, $D_x$ and $D_x'$ are the dispersion and its derivative, $\langle\;\rangle_s$ denotes averaging over the ring,

$$L_C = \ln\left(\frac{r_{max}}{r_{min}}\right), \quad r_{max} = \min\left(\sigma_x,\sigma_y,\sqrt{\frac{\theta_x^2+\theta_y^2}{Nr_p}\gamma^3\beta^2\sigma_x\sigma_y\sigma_s}\right), \quad r_{min} = \frac{2r_p}{\gamma^2\beta^2(\theta_x^2+\theta_y^2)} \qquad (8)$$

is the Coulomb logarithm ($L_C \approx 23$ for the case of the Tevatron). Lastly, the function

$$\Xi_s(x,y) \approx 1 + \frac{\sqrt{2}}{\pi}\ln\left(\frac{x^2+y^2}{2xy}\right) - 0.055\left(\frac{x^2-y^2}{x^2+y^2}\right)^2, \qquad (9)$$

approximates the exact result (obtained for Gaussian distribution) with accuracy better than a few percent, which is sufficiently good for all practical applications.

For small amplitudes, the rms bunch length growth rate due to RF noise is equal to

$$\left.\frac{d(\sigma_\phi^2)}{dt}\right|_{RF} = \pi \Omega_s^2 \left( P_\phi(\Omega_s) + \frac{1}{2}\sigma_\phi^2 P_A(2\Omega_s) \right). \quad (10)$$

Here $\sigma_\phi = 2\pi\sigma_s/\lambda_{RF}$ is the bunch length in radians, $\lambda_{RF}$ is the RF wavelength, and the spectral densities of the phase and amplitude noise are normalized as following

$$\overline{\delta\phi_{RF}^2} = \int_{-\infty}^{\infty} P_\phi(\omega)d\omega \quad , \quad \frac{\overline{\delta A_{RF}^2}}{A_{RF}^2} = \int_{-\infty}^{\infty} P_A(\omega)d\omega. \quad (11)$$

The effect of the RF noise on the Tevatron beam is dominated by the RF phase noise [6]. The main noise source is microphonics excited in the RF cavities due to the flow of cooling water. RF phase feedback suppresses this noise by ~30 dB to an acceptable level. Presently, the spectral density of the noise is about $P_{\phi f}(\Omega_s/2\pi) = 4\pi P_\phi(\Omega_s) \approx 5 \cdot 10^{-11}$ rad$^2$/Hz, which causes the bunch lengthening of about 2200 mrad$^2$/hour. This value is more than an order of magnitude smaller than the longitudinal emittance growth due to IBS at the beginning of a store with nominal proton intensity.

Besides the mechanisms described above, the Tevatron luminosity evolution model [7] takes into account the particle loss and the emittance growth due to collisions at IPs and collisions with residual gas atoms, as well the fact that the beam intensities, transverse and longitudinal emittances are changing during stores, thus, affecting the IBS diffusion rates and particle loss rates from the RF buckets. Figure 4 shows the bunch lengthening for proton and antiproton bunches calculated using the model (solid lines) in comparison with the measurements during a typical Tevatron HEP store (dashed line). Good agreement between the simulation and the observation for the protons indicates that the IBS is the main cause of longitudinal diffusion. For antiprotons, however, the beam-beam interaction with high intensity proton bunches [8] results in the loss of particles with large synchrotron oscillation amplitudes, slowing the bunch lengthening relative to that predicted by the IBS model. However since 2007, the brightness of the antiproton beam has been greatly increased that made large synchrotron oscillation protons more susceptible to the beam-beam effects, especially at the beginning of the HEP stores.

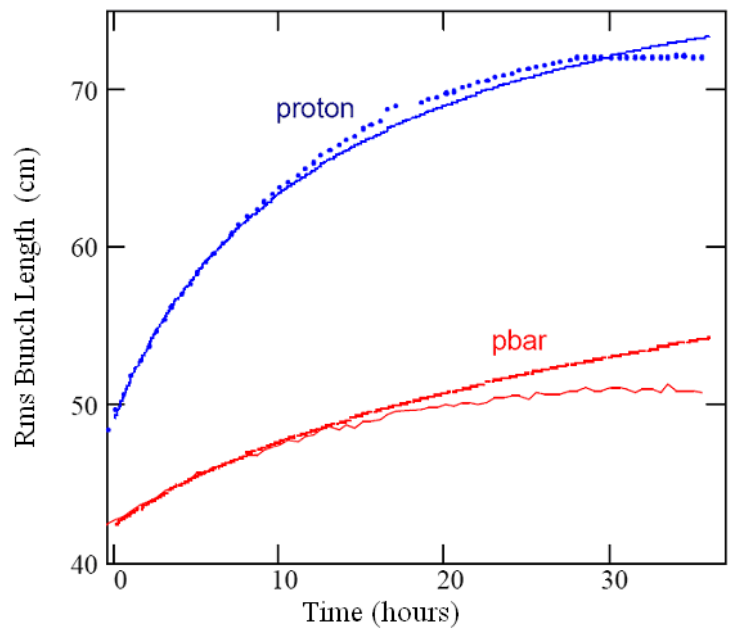

Figure 4: The simulated bunch lengthening effects compared with the measurements of a typical store (#3678).

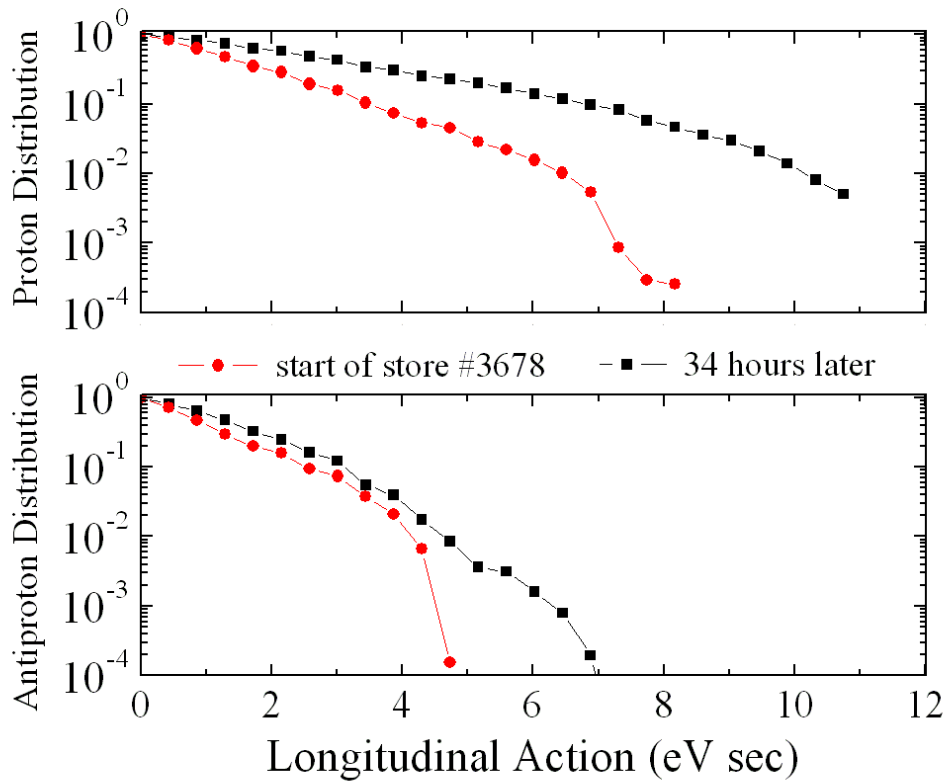

Figure 5. The measured dependence of the normalized longitudinal phase space density on the particle action for the protons (top) and antiproton (bottom) at the beginning (red) and at the end (black) of store #3678.

Figure 5 shows the evolution of the normalized longitudinal phase space density of the proton and antiproton beams in the Tevatron during a long store. At injection, both beams are contained at the 95% level within 4 eV s. During the store, the protons gradually diffuse out to the edge of the RF bucket at about 11 eV s where they can cross the edge and contribute to the uncaptured beam. The antiproton bunches, which are only about one seventh of the proton bunch intensity ($N_a=34e9$, $N_p=248e9$ per bunch at the start of the store), are effectively clipped by the beam-beam interaction and remain at essentially their initial longitudinal emittance. The IBS model allows us to calculate the longitudinal beam loss rate in a typical Tevatron store (Figure 6). The initial longitudinal loss rate is not equal to zero because of the Touschek effect. Later in the store, when more particles moves closer to the boundaries of the RF buckets through diffusion process, multiple IBS scattering starts to dominate over the single scattering effect. Note that for antiprotons, luminosity burning is the main loss contribution and the longitudinal loss due to IBS is much smaller than its total intensity loss rate.

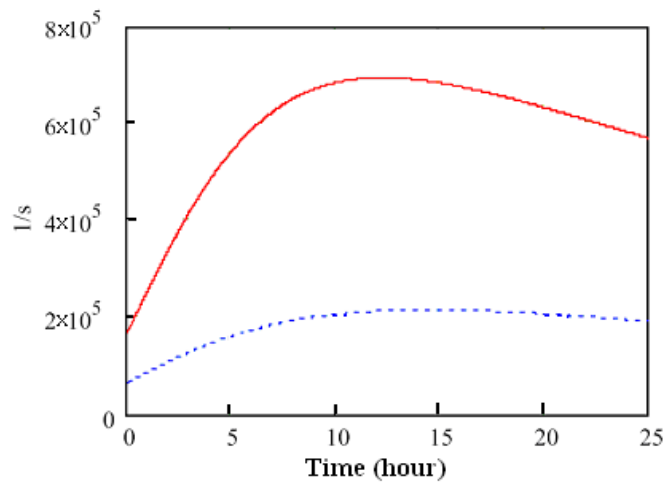

Figure 6: Calculated longitudinal beam loss rate in unit of particles per second for a typical store using the IBS model, the red curve is for the proton bunch and the blue dashed line is for the antiproton bunch.

The shape of the curve of the calculated longitudinal beam loss rate is in good qualitative agreement with the Tevatron observations. For example, Figure 7a shows the evolution of the total proton bunched beam intensity, proton loss rate, proton rms bunch length and the abort gap beam intensity during HEP store #5157. Bunch length and bunch intensity are reported from a wall current monitor (known as "Sampled Bunch Display" and briefly described in Ref. [9]). The loss rate is measured by gated scintillation counters on the CDF detector, which integrate over the time intervals corresponding to the abort gaps between the three proton bunch trains, while the simulation assumes for the whole Tevatron storage ring.

*2.3 Other mechanisms of uncaptured beam generation*

Large-amplitude beam-phase oscillations within the RF bucket due to instability or a sudden change of the RF bucket parameters (for example, an RF cavity tripping off) can result in large spills of particles into the uncaptured beam.

At the Tevatron injection energy of 150 GeV, large (~1 rad peak-to-peak) beam longitudinal dipole oscillations, termed "dancing bunches" [10] which are mainly caused by the coalescing process in the MI, are observed in high intensity beams and can persist for many minutes if not damped. These "dancing bunches" result in slow bunched-beam intensity loss and an increase in uncaptured beam which is lost at the start of acceleration. Another manifestation of the longitudinal impedances is the regular occurrence of large beam RF phase oscillations resulting in bunch lengthening. Such blowups again cause significant reduction of luminosity; increase beam losses, and accumulation of particles in the abort gaps. To counteract that, a longitudinal bunch-by-bunch damper was designed, built, installed and commissioned in the Tevatron in 2002 [2]. Since then, the damper has been in operation for every HEP store. It effectively suppresses both the "dancing bunches" and the single- and coupled-bunch instabilities. It was found that to be effective, the damper feedback loop gain *g* should vary slowly during the store in a fashion which tracks the proton bunch intensity and bunch length $g \sim N_p/\sigma_s$ [11].

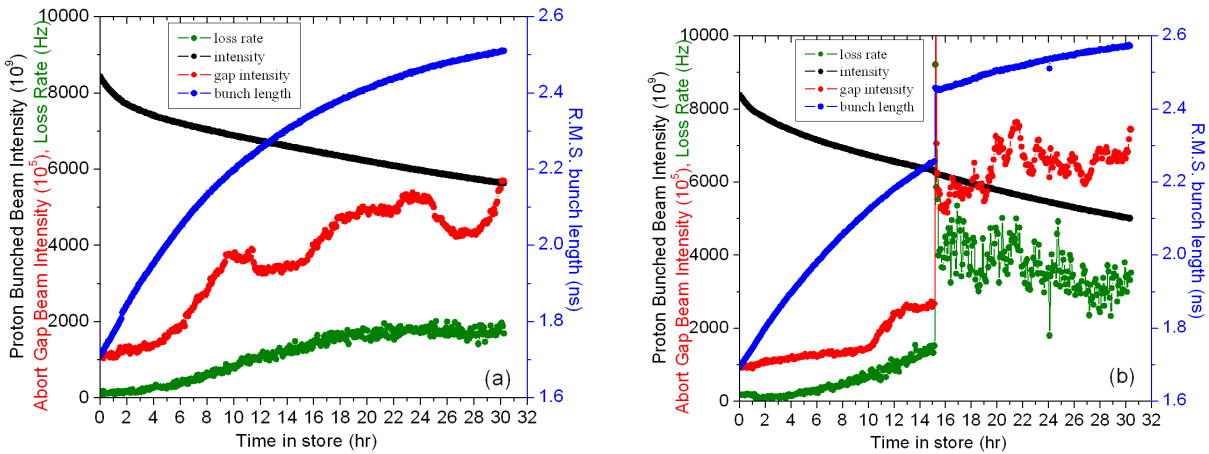

Figure 7. a) left – Decay of proton bunch intensity (black curve) and growth of its length (blue) as well as abort gap loss rate at CDF detector (red) and proton abort gap

intensity (green) in a typical HEP store (#5157) when the TEL was on; b) right – same for store #5136, in which one of 4 proton RF station tripped off in the middle of the store.

The Tevatron RF system consists of four drift tube cavities phased for acceleration of protons and four cavities phased for antiprotons. When one of the power amplifiers feeding these cavities trips off, the total RF voltage seen by the beam is decreased. That usually results in significant bunch lengthening, an instantaneous spill of particles into the uncaptured beam, and an increase of the uncaptured beam generation rate. The Tevatron protection system will immediately terminate HEP store if more than one RF station trips off in the same store because of the very high risk of damaging CDF or D0 electronics during an abort with an abnormally high intensity of uncaptured beam in the abort gaps. Figure 7b shows the beam parameters in store #5136 in which one of the four proton RF stations is tripped off about 16 hours after the beginning of the store. One can see a spike in abort gap losses and uncaptured beam intensity after the trip. The next sections describe the Tevatron uncaptured beam diagnostic tools and the Tevatron Electron Lens (TEL) operation as an abort gap beam remover.

## 3. UNCAPTURED BEAM DETECTION SYSTEMS

In the Tevatron, the uncaptured beam diagnostics are based on the detection of synchrotron radiation (SR) emitted by the particles in an SC dipole magnet. The method works only at 980 GeV where SR power is sufficiently large. The challenge in detecting the relatively small intensities associated with uncaptured beam is to avoid being blinded by the main particle bunches. This is accomplished with a gated photomultiplier tube (PMT) that observes the optical SR mainly originating from the dipole magnet edge. The system [12] is located in a short non-cryogenic section located between a full dipole and a half dipole where two moveable mirrors are positioned to intercept the light originating from the far edges of the adjacent dipoles. One mirror picks off the light from the protons, and the other mirror picks off the light generated by the antiprotons. The light exits the beam pipe through a quartz vacuum window and enters an optical box. Each optical box (see Figure 8) contains the PMT, optical attenuators, lens and intensified camera for producing transverse images of the beam.

The PMT is a modified Hamamatsu R5916U-50 (built with 3 stages of micro channel plates) with a maximum gain of ~$10^7$ and a minimum gating width of 5ns. It also has a large extinction ratio and no detectable sensitivity to light present before the gate is applied so that it will not be blinded by a preceding bunch. The shape and intensity of the magnetic field at the entrance to a dipole magnet results in roughly 25,000 (60,000) photons generated per 100nm bandwidth per $10^9$ protons

(antiprotons). The optical system efficiency reduces that amounts to 200 (500) PMT photoelectrons per $10^9$ protons (antiprotons). The data acquisition integrates the output of the PMT and averages over 1000 turns. Based on the stability of the PMT gain and the measurement-to-measurement variation, the detection sensitivity is estimated to be ~$10^6$ particles.

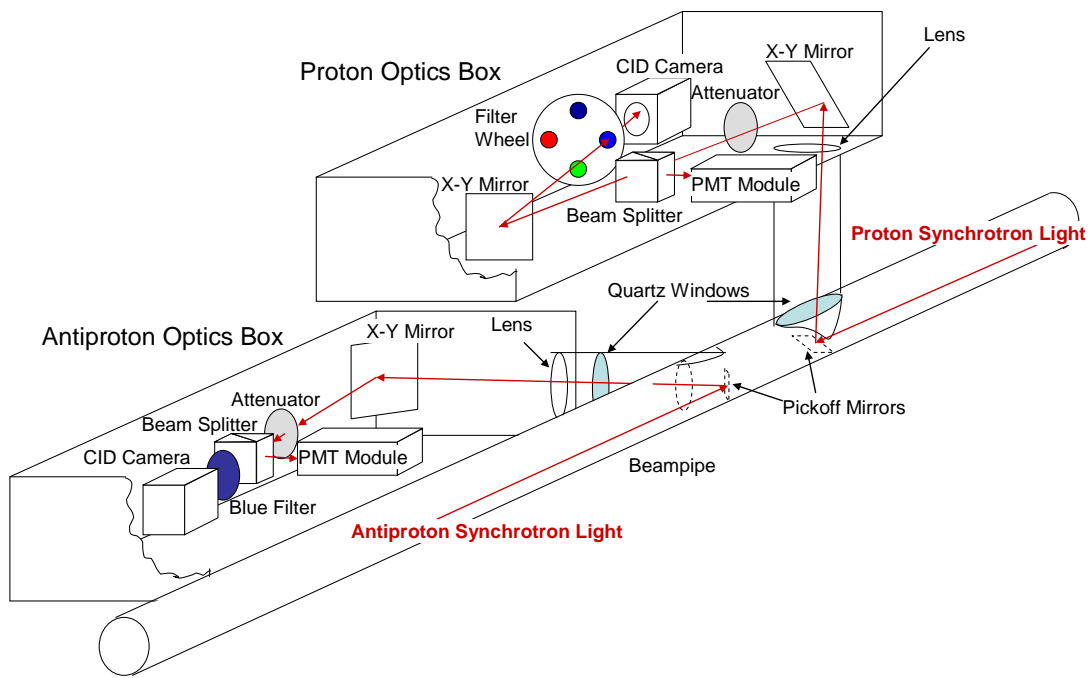

Figure 8: Diagram of optical collection system. There is one light box for the protons, and one for the antiprotons.

The calibration of the PMT can be accomplished in two ways. One method is to insert the optical attenuators and gate the PMT in coincidence with a bunch for which the intensity is known. This method relies heavily on the linearity of the PMT, but offers a simple technique for monitoring changes in the calibration. A second more reliable method utilizes the TEL and a DC current transformer measurement of the total beam current and is discussed below.

Figure 9 demonstrates the sensitivity of the PMT system. The data was taken by a digital oscilloscope to collect time information from single photon events. The micro-bunch structure in the abort gap coincides with RF buckets and is attributable to remnants from coalescing. And the high peaks around bunch 1 are the satellite bunches also left from the MI coalescing process.

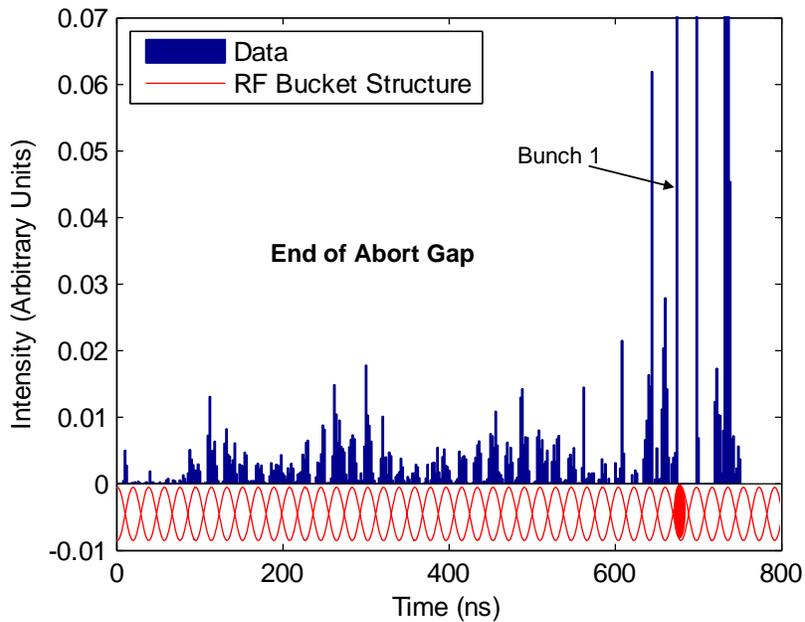

Figure 9: Structure of the beam in the tail end of the abort gap. Blue bars represent proton intensity; red curves are given for reference and show the Tevatron RF buckets.

A comparison of bunched beam intensity (measured by the Sampled Bunch Display or Fast Bunch Integrator systems [11]) with total beam intensity measured by a DCCT gives in principle an alternative estimate of the uncaptured beam intensity. Unfortunately, the systematic errors of the bunched beam intensity monitors are too large (~$10^{11}$ protons for a total beam intensity of about $10^{13}$) [11]. Therefore this method is used only at 150 GeV where the synchrotron light diagnostic is inoperable.

## 4. UNCAPTURED BEAM REMOVAL

As explained in the introduction, the presence of the uncaptured beam is very dangerous for the collider elements and the high-energy physics particle detectors CDF and D0. A number of ideas have been proposed for elimination of the uncaptured beam in the Tevatron. The Tevatron Electron Lenses have been found to be the most effective [4]. As explained in this section, the advantages of the TELs are two-fold: i) an electron beam is in close proximity to proton or antiproton orbits and generates a very strong transverse kick; ii) the TELs possess short rise and fall times (~100 ns), so they can be easily adjusted to operate in a variety of different pulsing schemes. Another uncaptured beam removal method tested during machine studies was a transverse strip line kicker operating with a narrow noise bandwidth. The kicker signal was timed into the abort gap to diffuse uncaptured beam particles transversely. With the noise power limited by a 300 W amplifier, that method was found significantly less effective than using the TELs [13]. Abort gap cleaning by very strong kicker magnets worked effectively in HERA and SPS[14].

*4.1 TEL as the uncaptured beam cleaner*

The Tevatron Electron Lenses #1 and #2 were installed in the Tevatron in 2001 and 2006, correspondingly, for compensation of beam-beam effects [15]. In 2002, it was found that TEL-1 can very effectively remove uncaptured beam protons if timed into the abort gap and operated in a resonant excitation regime [4]. TEL-2 is also able to function as an abort gap cleaner.

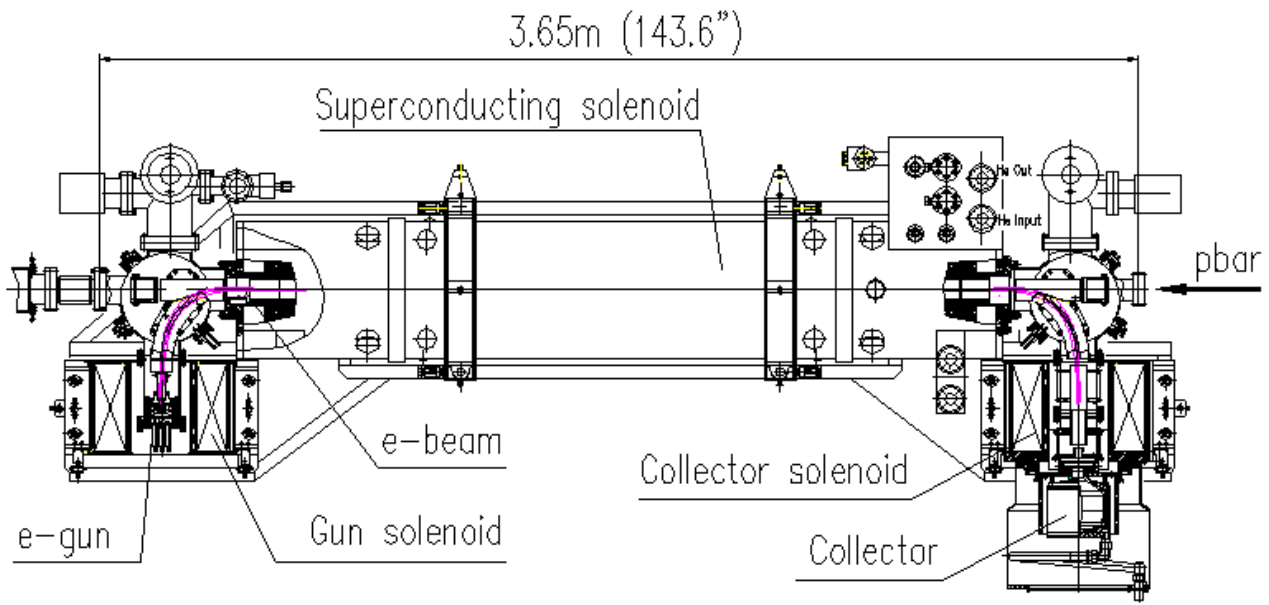

Figure 10: Layout and main components of the first Tevatron Electron Lens.

Figure 10 shows the layout of the TEL-1. The magnetic system of the TEL consists of a 65 kG superconducting (SC) main solenoid, four 8 kG and two 2 kG SC dipole correctors in the same cryostat, and conventional 4 kG gun and collector solenoids. The TEL cryostat is part of the Tevatron magnet string cooling system. A strong Π-shaped magnetic field is needed to guide the 10 kV electron beam from the electron gun through the interaction region, where electrons interact with high-energy (anti)protons, to the collector. The low-energy electron beam of about 4 mm in diameter is strongly magnetized and follows the magnetic field lines. SC dipole correctors allow precise steering, in position and angle, of the electron beam onto the beams circulating in the Tevatron.

To operate the TEL as the abort gap uncaptured beam remover, the electron beam pulse is synchronized to the abort gap and positioned near the proton beam orbit. Electric and magnetic forces due to the electron space charge produce a radial kick on high-energy protons depending on the separation $d$:

$$\Delta\theta = \mp \frac{1 \pm \beta_e}{\beta_e} \cdot \frac{2 J_e L_e r_p}{e \cdot c \cdot \gamma_p} \cdot \begin{cases} \dfrac{d}{a}, & d < a \\ \dfrac{a}{d}, & d > a \end{cases} \qquad (12)$$

where the sign reflects repulsion for antiprotons and attraction for protons, $\beta_e = v_e/c$ is the electron beam velocity, $J_e$ and $L_e$ are the electron beam current and the interaction length, $a$ is the electron beam radius, $r_p$ is the classical proton radius, and $\gamma_p = 1044$ is the relativistic Lorentz factor for 980 GeV (anti)protons. The factor $1 \pm \beta_e$ reflects the fact that the contribution of the magnetic force is $\beta_e$ times the electric force contribution and depends on the direction of the electron velocity.

For a typical peak current of about 0.6 A and 5 kV electrons 5 mm away from the protons, the estimated kick is about 0.07 μrad. When the pulsing frequency of the

TEL is near the proton beam resonant frequency, this beam-beam kick resonantly excites the betatron oscillations of the beam particles.

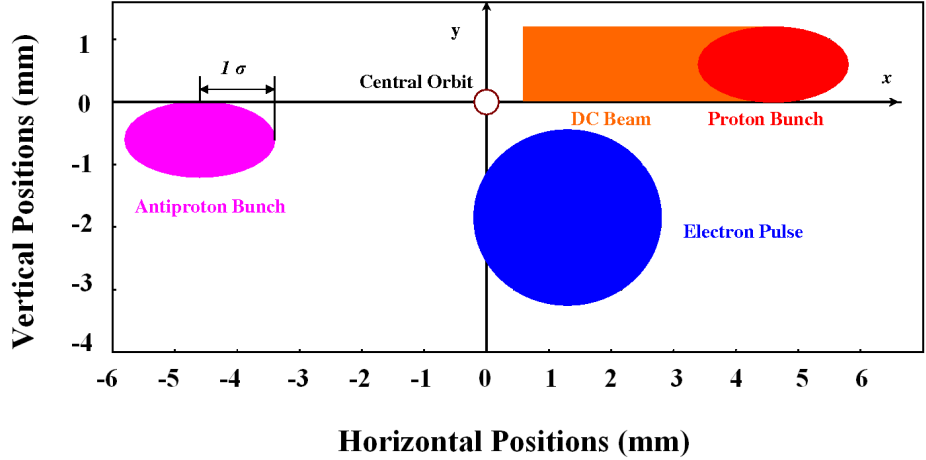

Figure 11: The relative positions of the proton, antiproton and electron beam during uncaptured beam removal operation.

In uncaptured beam removal operation, the TEL electron beam is placed 2-3 mm away from the proton beam orbit horizontally and about 1 mm down vertically as schematically depicted in Figure 11. For normal Tevatron operation, the fractional part of the tunes are $Q_x= 0.583$ and $Q_y= 0.579$ for horizontal and vertical planes respectively. This tunes are placed between the strong resonances at $4/7 \approx 0.5714$ and $3/5=0.6$. When an uncaptured particle loses energy due to synchrotron radiation, its horizontal orbit is changed proportionally to the lattice dispersion $x=D_x(dP/P)$ and its betatron tunes are changed due to the lattice chromaticity $C_{x,y}=dQ_{x,y}/(dP/P)$:

$$Q_{x,y} = Q^0_{x,y} + C_{x,y}\left(\frac{dP}{P}\right) + \Delta Q_{x,y}(x^2) \quad , \quad (13)$$

where the third term accounts for slight tune changes due to nonlinear magnetic fields. Typical operational chromaticities of the Tevatron at 980 GeV are set to $C_{x,y} \approx +10$, so the tune decreases with the energy loss. As the tune, driven by the TEL, approaches one of the resonant lines, the amplitude of the particle betatron oscillations grows, eventually exceeding a few millimeters until the particle is intercepted by the collimators. Figure 12 presents one set of the simulation results of the particle amplitude driven by the TEL to the vicinity of the $4/7^{th}$ resonance. The maximum amplitude is determined by the nonlinearity of the force due to the electron beam and nonlinearity of the machine. Note that without the TEL, a particle would still be intercepted by a horizontal collimator after its orbit moved about 3 mm inward due to SR. The TEL simply drives it more quickly, preventing the accumulation of DC-beam particles.

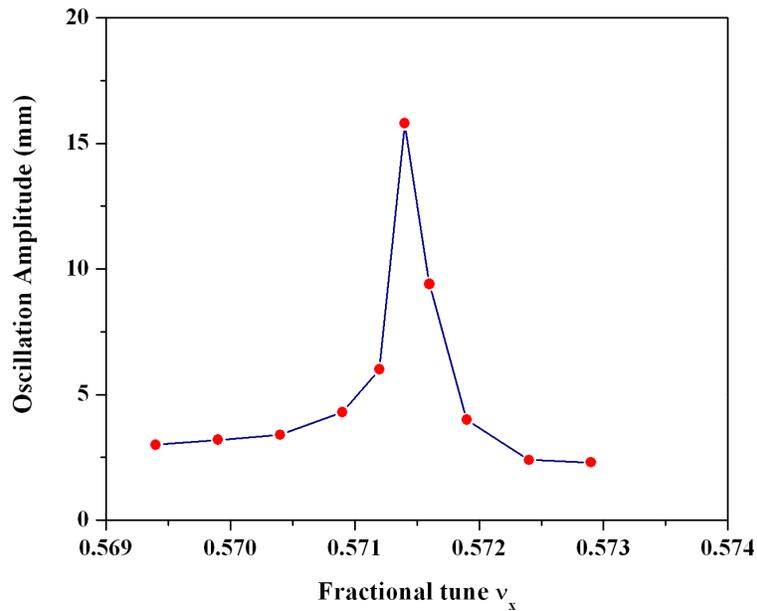

Figure 12: Betatron oscillation amplitude of the particles driven by the TEL in vicinity of $Q=4/7$th resonance line (simulations).

The electron beam pulsing scheme is demonstrated in Figure 13, where the green oscilloscope trace is the signal from the TEL Beam Position Monitor (BPM) pickup electrode and the blue trace is the total electron current. In the BPM signal, one can see three negative pulses representing the electron beam pulses in the 3 abort gaps whereas the 36 positive pulses are the proton bunch signals with the small negative adjacent antiproton bunch signals. The intensity of the antiproton bunches was 10 times less than that of the proton bunches at the end of that particular store, so they appear only as very small spikes near the large proton bunches. During a typical HEP store, the train of three electron pulses is generated every 7[th] turn for the purpose of excitation of the 4/7 resonance for the most effective removal of the uncaptured beam proton particles. The electron pulse width is about 1 µs and the peak amplitude is about 250 mA.

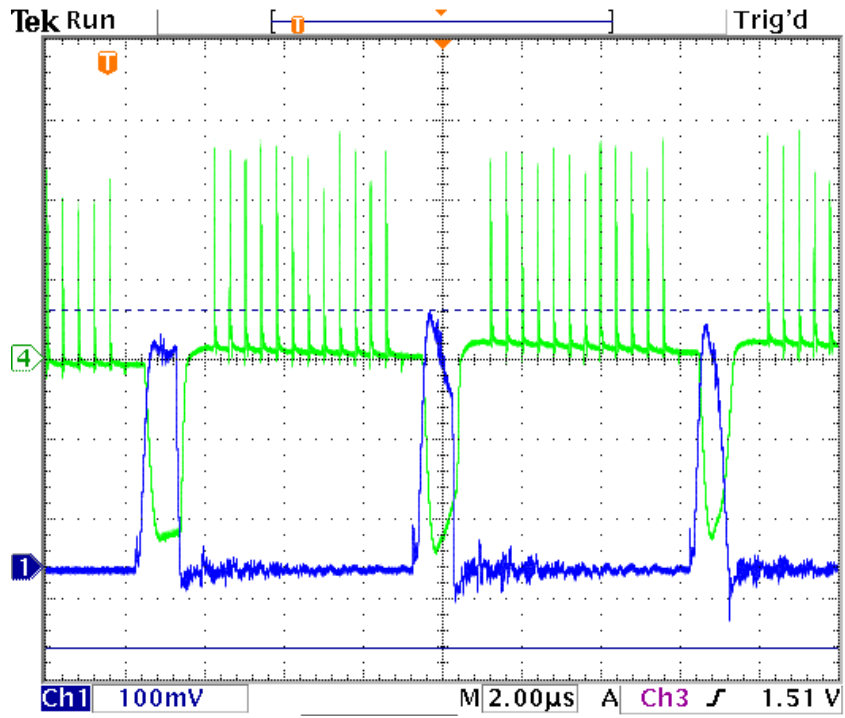

Figure 13: Scope traces of the electron beam pulses (blue) and the TEL BPM signal showing electron, proton and antiproton bunches. One division of the horizontal axis is 2 microseconds. About one Tevatron revolution period is shown here.

The uncaptured beam removal process is demonstrated in an experiment in which the TEL was turned off for about 40 min and then turned on again as showing in Figure 14. The blue trace is the total bunched proton beam intensity measured by the Fast Bunch Integrator [9]; the red trace is the average electron current measured at the TEL electron collector; the green trace is the total number of particles in the Tevatron as measured by DCCT [9]; and the cyan trace is the abort gap proton beam loss rate measured by the CDF detector counters.

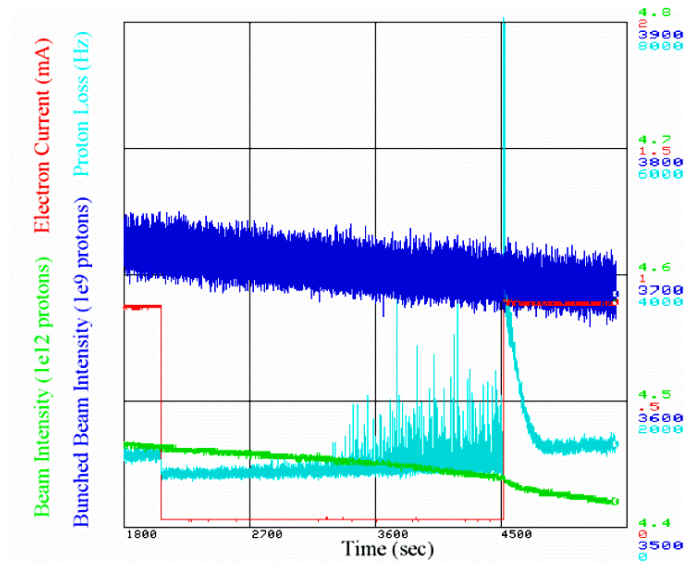

Figure 14: Uncaptured beam accumulation and removal by TEL: the electron current was turned off and turned back on 40 min later again.

After the TEL was turned off, the abort gap loss rate reduced by about 20% but then started to grow. After about 20 min, the first spikes in losses started to appear and grow higher. Notably, the bunched beam intensity (blue line) loss rate did not change, so the rate of particles escaping from the RF buckets was about constant. As soon as the TEL was turned on, a very large increase of the abort gap losses and reduction of the total uncaptured beam intensity could clearly be seen in Figure 14. About $15 \times 10^9$ particles of the uncaptured beam in the abort gaps were removed by the TEL in about $\tau_{TEL}=3$ minutes and the abort gap loss rate went back to a smooth equilibrium baseline.

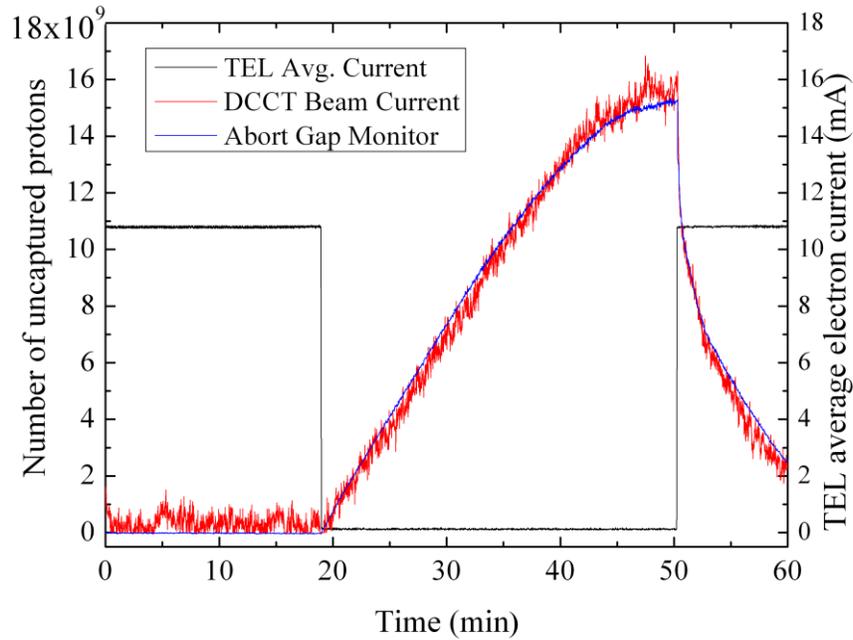

Figure 15: Uncaptured beam accumulation and removal by the TEL. The black line represents the average electron current of the TEL; the blue line is the uncaptured beam estimated from the DCCT measurement; the red line is uncaptured beam in the abort gap measured by the AGM.

The calibration of the abort gap monitor (AGM) has been performed using the TEL as presented in Figure 15. The TEL was turned off during a store (average electron current is shown in black) at about $t = 20$ min. Accumulation of the uncaptured beam started immediately and can be measured as an excess of the total uncaptured beam current with respect to its usual decay. The blue line in Figure 15 shows the excess measured by the Tevatron DCCT, $\delta N_{DCCT}(t) = N_{TEL\ on}(t) - N_{decay\ fit\ TEL\ off}(t)$. The abort gap uncaptured beam intensity measured by the AGM (red line) and the DCCT excess grow for about 30 minutes before reaching saturation at intensity of about $16 \times 10^9$ protons. Then the TEL was turned on resulting in the quick removal of the accumulated uncaptured beam from the abort gaps. This method of calibration of the AGM with respect to DCCT interferes with the collider operation resulting in higher losses (see Figure 14 above and discussion), so this operation is performed only when required. The AGM is used for the routine monitoring of the uncaptured beam. The typical rms error of the uncaptured beam intensity measurement is about $0.01 \cdot 10^9$ protons for the AGM, and some $0.3 \cdot 10^9$ protons for the DCCT.

The amount of the uncaptured beam is determined by the rate of its generation and the removal time $\tau$:

$$N_{DC} = \left(\frac{dN_{bunched}}{dt}\right) \times \tau \qquad (14)$$

The characteristic time needed for a 980 GeV particle to lose enough energy due to SR is about $\tau_{SR}=20$ minutes, so the TEL reduces the uncaptured beam population by about one order of magnitude.

At injection energy, the synchrotron radiation of protons is negligible, so the TEL is the only means to control uncaptured beam. As noted above, one of the TELs is used routinely in the Tevatron operation for the purpose of uncaptured beam removal at 150 GeV and 980 GeV. In 2007, typical antiproton intensity has been increased to about a third of the proton one, and therefore the antiproton uncaptured beam accumulation has started to pose an operational threat. An antiproton AGM, similar to the proton one, has been built and installed. By proper placement of the TEL electron beam between the proton beam and the antiproton beam (illustrated in Figure 11), we are able to remove effectively both un-captured protons and un-captured antiprotons. In addition, we have explored the effectiveness of the uncaptured beam removal at several resonant excitation frequencies. For that, we pulsed the TEL every $2^{nd}$, $3^{rd}$, $4^{th}$, $5^{th}$, $6^{th}$ and $7^{th}$ turn. Reduction of the uncaptured beam intensity was observed at all of them, though usually the most effective were the every $7^{th}$ turn pulsing when the Tevatron betatron tunes are close (slightly above) to $Q_{x,y}=4/7=0.571$ or every $6^{th}$ turn pulsing when tunes are closer to $Q_{x,y}=7/12=0.583$.

## 5. CONCLUSIONS

Uncaptured beam has been found to be very dangerous for Tevatron operation in the Collider Run II. We identified the main mechanisms of uncaptured beam generation, namely, the intrabeam scattering, RF noise, longitudinal instabilities and RF cavity trips. Sensitive uncaptured beam diagnostics have been developed on the basis of the synchrotron light monitors. The uncaptured beam intensity is controlled by using the Tevatron Electron Lenses for the removal of uncaptured particles. The TEL electron beam is synchronized with abort gaps and resonantly excites betatron oscillations of the (anti)protons which are then lost on the tightest beam aperture (collimators). The TELs smoothly remove the uncaptured beam from the abort gap within minutes. Experience with the uncaptured beam in the Tevatron, as well as in other hadron colliders such as HERA and RHIC [14], provides valuable input for uncaptured beam control in CERN's Large Hadron Collider [16].

We would like to thank J.Annala, A.Burov, B.Drendel, B.Hanna, T.Khabibilin, D.McGinnis, E.McCrory, R.Moore, J.Morgan, J.Reid, G.Romanov, J.Steimel, D.Still, C.Y.Tan and F.Zimmermann for valuable input, assistance during beam studies and useful discussions on the subject.

Fermilab is operated by Fermi Research Alliance Ltd. under Contract No. DE-AC02-76CH03000 with the United States Department of Energy.